\documentclass[twocolumn]{aastex631}
\AtBeginDocument{%
}

\usepackage{xcolor}
\usepackage{amsmath}
\usepackage{multirow}

\def\flunits{\mathrm{erg\; cm^{-2}\; s^{-1}\; \AA^{-1}}}
\def\cycleday{${\rm d}^{-1}$}
\def\kms{km~s$^{-1}$}

\def\ergsec{erg~s$^{-1}$}

\begin{document}

\title{V844~Herculis: An Exceptional Intermediate Polar Hiding in an Ordinary Dwarf Nova}

\author[0009-0008-6389-9398]{Anousha Greiveldinger}
\affiliation{Department of Physics and Astronomy, University of Notre Dame, Notre Dame, IN 46556, USA}
\affiliation{Department of Physics, University of California, Berkeley, Berkeley, CA 94720, USA}
\affiliation{Physics Division, Lawrence Berkeley National Laboratory, 1 Cyclotron Road, Berkeley, CA, 94720}

\author[0000-0003-4069-2817]{Peter Garnavich}
\affiliation{Department of Physics and Astronomy, University of Notre Dame, Notre Dame, IN 46556, USA}

\author[0000-0001-7746-5795]{Colin Littlefield}
\affiliation{Department of Physics and Astronomy, University of Notre Dame, Notre Dame, IN 46556, USA}
\affiliation{Bay Area Environmental Research Institute, Moffett Field, CA 94035 USA}

\author[0000-0001-6894-6044]{Mark R. Kennedy}
\affiliation{School of Physics, University College Cork, Cork, Ireland}
\affiliation{Jodrell Bank Centre for Astrophysics, Department of Physics and Astronomy, The University of Manchester, Manchester M13 9PL, United Kingom}

\author[0000-0003-4373-7777]{Paula Szkody}
\affiliation{Department of Astronomy, University of Washington, Seattle, WA 98195, USA}

\author[0009-0003-0467-4440]{Jordan Tweddale}
\affiliation{Department of Physics and Astronomy, University of Notre Dame, Notre Dame, IN 46556, USA}

\author{Mark Whittle}
\affiliation{Department of Astronomy, University of Virginia, 530 McCormick Road, Charlottesville, VA 22903, USA}

\author{Te\'ofilo Arranz}
\affiliation{Observatorio Astron\'omico ``Las Pegueras", Navas De Oro, Segovia, Spain}

\author{Donald Collins}
\affiliation{Warren Williams College, 701 Warren Wilson Road, Swannanoa, NC 28778}

\author{James DeYoung}
\affiliation{AAVSO, 185 Alewife Brook Parkway, Suite 410, Cambridge, MA 02138, USA}

\author{Sjoerd Dufoer}
\affiliation{Vereniging Voor Sterrenkunde (VVS), Oostmeers 122 C, 8000 Brugge, Belgium}

\author[0000-0002-8908-0785]{Charles Galdies}
\affiliation{Institute of Earth Systems, University of Malta, Malta}

\author{Jean-Francois Gout}
\affiliation{AAVSO, 185 Alewife Brook Parkway, Suite 410, Cambridge, MA 02138, USA}

\author[0000-0003-0125-8700]{Franz-Josef Hambsch}
\affiliation{Vereniging Voor Sterrenkunde (VVS), Oostmeers 122 C, 8000 Brugge, Belgium}
\affiliation{Groupe Europ\'een d’Observations Stellaires (GEOS), 23 Parc de Levesville, 28300 Bailleau l’Ev\^eque, France}
\affiliation{Bundesdeutsche Arbeitsgemeinschaft f\"ur Ver\"anderliche Sterne e.V. (BAV), Munsterdamm 90, 12169 Berlin, Germany}

\author{Emmanuel Kardasis}
\affiliation{Pelagia-Eleni Observatory, Athens, Greece}

\author{David Messier}
\affiliation{CBA Norwich, 35 Sergeants Way, Lisbon, CT 06351, USA}

\author[0000-0001-6249-6062]{Arto Oksanen}
\affiliation{AAVSO, 185 Alewife Brook Parkway, Suite 410, Cambridge, MA 02138, USA}

\author{Gary Poyner}
\affiliation{BAA Variable Star Section, 67 Ellerton Road, Kingstanding, Birmingham B44 0QE, UK}

\author{Richard Sabo}
\affiliation{CBA-Montana, 1344 Post Dr., Bozeman, MT 59715}

\begin{abstract}

We present time-series spectroscopy of the SU~UMa-type dwarf nova V844~Her during quiescence. We discover high-velocity Balmer and He~I emission features modulated at a period of  29.3$\pm0.1$~minutes. The spectroscopic periodicity is identical to the photometric period detected only during superoutbursts. Interpreting this as the result of magnetic accretion onto the white dwarf, we identify V844~Her as an intermediate polar (IP). The spin period of the white dwarf is most likely to be 29.3~minutes. In a few IP systems, optical modulation is seen at twice the spin frequency, so we can not rule out that the spin period is 58.6~minutes. The orbital period of V844~Her is only 1.31~hr, meaning that the system has the shortest binary period and one of the smallest orbit-to-spin ratios of the confirmed IPs. We present extensive optical, ultraviolet, and X-ray observations of V844~Her during the 2023 superoutburst. Compared with quiescence, the XRT count rate decreases during superoutburst. While the enhanced mass transfer during the superoutburst likely increases X-ray production, we propose that the drop in the count rate comes from the loss of soft X-rays due to an additional column from the outbursting disk.

\end{abstract}


\section{Introduction}

Cataclysmic variables (CV) are binary star systems consisting of an accreting white dwarf (WD) and a late-type secondary star. In CVs, the cool secondary star donates gas to the WD and the mass generally sheds its angular momentum through the formation of an accretion disk. A significant magnetic field on the WD can strongly influence the accretion process. A surface field of more than about 10~MG can prevent a disk from forming \citep{hellier14}. Such a system is classified as a polar, and the field is often sufficiently strong to lock the WD spin to the binary orbital period. An intermediate polar (IP) is a system in which the WD is moderately magnetic so that a partial disk can form and the WD may spin asynchronously. The WDs in IPs generally have surface fields less than 10~MG, but the lower limit on an IP field has not been determined. The asynchronous WD spin in IPs can result in a wide range of photometric periodicities, including those at the orbital frequency, spin frequency, beat frequency (difference between the spin and orbital frequencies), and their harmonics. In IPs and polars, accretion channeled along the WD magnetic field lines often results in significant X-ray emission from a shock formed near the surface and heating around the magnetic poles.

V844~Her has been classified as an SU~UMa-type dwarf nova (DN) due to evidence of long and bright superoutbursts.  While many CVs display ordinary DN outbursts that arise from the thermal instability of their accretion disks, superoutbursts occur in short-period CVs and include dynamical disk instabilities. V844~Her has an orbital period of only 78.69~min \citep{thorstensen02}, which is near the period minimum for hydrogen-rich CVs \citep{knigge11}. The measured geometric distance is $305\pm 5$~pc \citep{Bailer-Jones21}. V844~Her has not been suspected of being an IP, however, \citet{anousha23} identified a new 29-minute periodicity in its light curve based on data from NASA's Transiting Exoplanet Sky Survey \citep[TESS;][]{ricker15} mission. This new oscillation was detected by TESS only during a superoutburst. It appears unrelated to the system's orbital frequency or super-hump oscillations. \citet{anousha23} suggested several possible sources for the new periodicity, with an asynchronously spinning WD possessing a significant magnetic field as the most plausible explanation.

Here, we present extensive optical, ultraviolet (UV), and X-ray observations of V844~Her during a recent superoutburst. Further, we analyze new time-resolved optical spectra of V844~Her taken during quiescence.

\section{Data}

Prompted by the discovery of a new photometric periodicity during a superoutburst of V844~Her \citep{anousha23}, we requested a monitoring campaign by the American Association of Variable Star Observers (AAVSO) to trigger intense observations early in a new superoutburst. On 2023 October 1.00 (UT), J. DeYoung noted that V844~Her was 0.7~mag brighter than its typical quiescent state. By October 1.80 UT, G. Poyner noted a visual brightness of 12.6 mag, confirming the onset of a large amplitude outburst and likely a superoutburst. 

At the initiation of the outburst, we requested XRT and UVOT observations of V844~Her with the Neil Gehrels Swift Observatory \citep[Swift hereafter;][]{swift04}.

\subsection{Photometry} 

\subsubsection{SLKT Photometry}

The Sarah L. Krizmanich Telescope (SLKT) began time-series photometric observations of V844~Her soon after the onset of the outburst. The SLKT is located on the University of Notre Dame campus and features a 0.8-m (32-in diameter) primary mirror. We employed an SBIG camera with a CMOS detector allowing exposure times of 3~s with only 0.8~s of overhead between images. The data were obtained unfiltered. Processing consisted of subtracting 3~s dark exposures and dividing by a flat field image created from twilight sky exposures. The brightness of V844~Her was estimated using aperture photometry on the target and nearby reference stars. 

\subsubsection{AAVSO}
\begin{figure}
    \centering
    \includegraphics[width=\columnwidth]{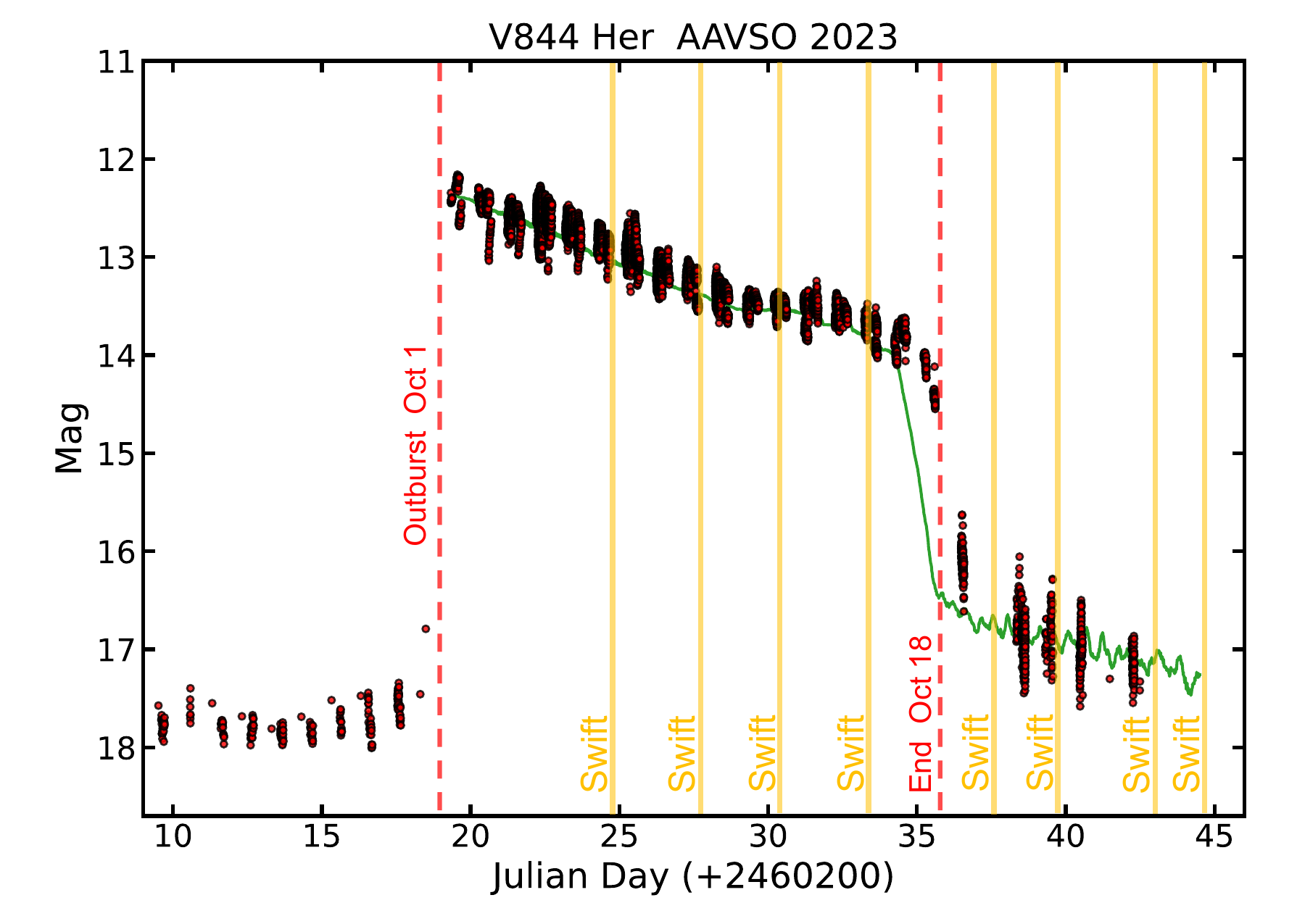}
    \caption{The AAVSO light curve of V844~Her during the 2023 superoutburst. Circles show the approximately 24000 individual photometric measurements. The solid line is the median-smoothed TESS light curve from the 2020 outburst, shifted in time to match the onset of the 2023 outburst. The dashed vertical bar on the far left marks the beginning of the outburst on 2023 October 1 (UT). The right dashed vertical bar marks the end of the outburst on 2023 October 18. The remaining vertical bars indicate visits by the Swift satellite collecting UV and X-ray data. 
    \label{aavso}}
\end{figure}

Contributors to the American Association of Variable Star Observers (AAVSO) International Database \footnote{https://www.aavso.org/LCGv2/} observed V844~Her throughout the duration of its superoutburst in October 2023. Data from the AAVSO beginning late September and throughout October is shown in Figure \ref{aavso}. Observations were generally obtained unfiltered and calibrated to the $V$-band.

\subsubsection{Swift}

V844~Her was observed by the Swift satellite on 8 visits over the 25 days after the start of the October~1 superoutburst using the UltraViolet and Optical Telescope (UVOT), and the X-ray Telescope (XRT). Archival observations from 2016, when V844~Her was in quiescence, were included in the analysis. Information about each Swift visit is given in Table \ref{swiftvis}.

We analyzed data from the UVOT \citep{Roming05} using the Fv FITS viewer \citep{Pence12} software. We filtered observations by offset and chose those in which V844~Her was closest to the center of the image, limiting the effects of vignetting and other variations across the detector. Using the ImageProbe feature, we extracted the total counts within an aperture centered on V844~Her for each instrument. We converted the measured count rates into a flux density using factors given in \citet{poole08} Table 9. We calculated error bars by dividing the square root of the counts by exposure time, then propagating the conversion factor errors. 

For the XRT \citep{Burrows05}, we used count rates from the Living Catalog \citep{evans23} that have made corrections for vignetting, detector, and exposure length.

The data from 2016-01-17 were used to calculate the quiescence flux levels for both the XRT and UVOT.

\begin{deluxetable*}{lcccccc}
\centering
\tablecaption{Log of Swift XRT and UVOT Observations   \label{swiftvis}}
\tablehead{
\colhead{$\textrm{Datset ID}$} & \colhead{$\textrm{Start Date}$} & \colhead{$\textrm{Start Date}$} & \colhead{$\textrm{XRT Exposure Length}$} & \colhead{$\textrm{XRT Count Rate}$} & \colhead{$\textrm{UVOT Flux Density}$} & \colhead{$\textrm{UVOT Filter}$} \\
\colhead{} & \colhead{$\textrm{(UT Date)}$} & \colhead{$\textrm{(JD +2460200)}$} &  \colhead{$\textrm{(ks)}$} & \colhead{$\textrm{(10$^{-2}$~s$^{-1}$)}$} & \colhead{$\textrm{(10$^{-16}$}$ $\flunits$ )} & \colhead{} }
\startdata
00045736001$^{a}$ & \multirow{2}{*}{2016-01-17} & \multirow{2}{*}{$-2795.42$} & \multirow{2}{*}{5.1} & \multirow{2}{*}{4.5$^{+0.3}_{-0.3}$} & - & - \\
00045736002 & & & & & 5.204$\pm$0.3 & $\textrm{u}$ \\
00045736003 & 2023-10-07 & 24.9 & 1.9 & 1.8$^{+0.4}_{-0.3}$ & 206.5$\pm$19.3 & \textrm{u}\\
00045736004 & 2023-10-10 & 27.7 & 0.90 & 1.5$^{+0.5}_{-0.4}$ &  398.4$\pm$19.5 & $\textrm{uvw1}$\\
00045736005 & 2023-10-13 & 30.6 & 2.1 & 1.8$^{+0.4}_{-0.3}$ & 471.9$\pm$69.2 & $\textrm{uvm2}$\\
00045736006 & 2023-10-16 & 33.5 & 2.0 & 1.7$^{+0.3}_{-0.3}$ & 460.5$\pm$49.1 & $\textrm{uvw2}$\\
00045736007 & 2023-10-19 & 37.3 & 1.2 & 0.5$^{+0.3}_{-0.2}$ & 27.26$\pm$2.55 & $\textrm{u}$\\
00045736008$^{b}$ & 2023-10-22 & 39.6 & 0.62 & 0.7$^{+0.7}_{-0.4}$ & 22.10$\pm$1.15 & $\textrm{uvw1}$\\
00045736009 & 2023-10-25 & 43.1 & 1.5 & 3.5$^{+0.6}_{-0.5}$ & 18.49$\pm$2.73 & $\textrm{uvm2}$\\
00045736010 & 2023-10-27 & 44.6 & 2.3 & 3.3$^{+0.4}_{-0.4}$ & 6.037$\pm$0.57 & $\textrm{u}$ \\
\enddata
\tablenotetext{a}{For the UVOT data, only ObsID 00045736002 was used, while for the XRT data, ObsID 00045736001 and 00045736002 were combined.}
\tablenotetext{b}{The Living Catalog \citep{evans23} lists this observation as a ``non-detection''.}
\end{deluxetable*}

\begin{figure}
    \centering
    \includegraphics[width=\columnwidth]{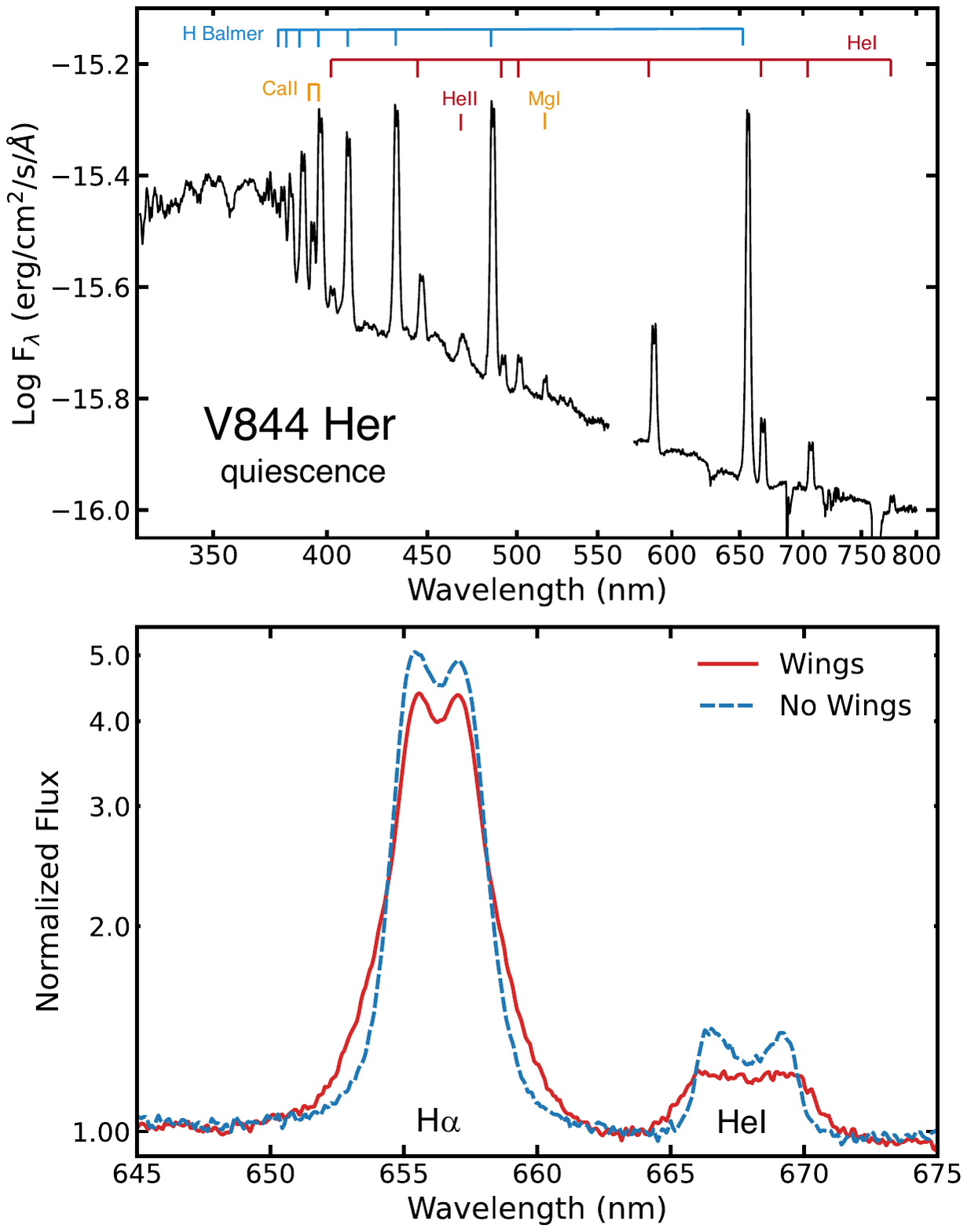}
    \caption{{\bf Top:} The average of all LBT spectra showing strong Balmer and neutral helium emission lines on a blue continuum. A broad but rather weak ionized helium line at 468.6~nm is detected. The gap in the data at 560~nm is due to the dichroic that divides the light between the MODS red and blue spectrographs. {\bf Bottom:} The average of 12 spectra showing high velocity emission wings (solid red line) compared with the average of the narrowest 10 spectra (dashed blue line). When the Balmer lines are broad, the peak flux is lower than when the lines are narrow, leaving the total flux nearly constant.
    \label{ave_spec}}
\end{figure}

\subsection{Optical Spectroscopy}

Time-series spectroscopy of V844~Her was obtained with the Multi-Object Dual Spectrograph \citep[MODS;][]{pogge12} mounted on the Large Binocular Telescope (LBT). The observations were performed on 2023 June 17 (UT) when V844~Her was in its quiescent state. MODS2 recorded 25 spectra using its red channel covering 565~nm to 1020~nm and 23 spectra on its blue side covering 320~nm to 560~nm. Exposure times were 180~s with 68~s between exposures for overheads. After 30~minutes of collecting data, the LBT was shutdown due to high winds. The time-series was resumed after the hour closure and 70~minutes of additional data were obtained. 

The spectral images were extracted and wavelength calibrated using Neon and Argon emission lamps. Over the time series, there was flexture in the spectrograph that was removed by small shifts in the wavelength solutions based on the airglow sky lines.  The spectra were flux calibrated using spectra of the standard star BD+33~2642 obtained on the same night as those for V844~Her. The resulting average spectrum is show in Figure~\ref{ave_spec}.

\section{Analysis of the Superoutburst}

\subsection{Optical Variability}

From the AAVSO light curve, we estimate the start of the outburst to be JD~2460219.0 (October 1, 2023 12:00 UT), as it was 0.7~mag brighter than its typical quiescent state on JD~2460218.5 and by 2460219.3 it had reached 12.4~mag.   ATLAS \footnote{``Asteroid Terrestrial-impact Last Alert System'', \citet{tonry18}} caught the star on its initial brightening with $o$-band estimates of 15.35 and 15.05~mag on JD~2460218.71 and 2460218.74 

The outburst began to steeply fade on JD~2460235.5, ending the bright plateau phase of the outburst 16.5 days after it began. V844~Her continued to slowly fade in the optical band through the last AAVSO time series on 2460242. An ATLAS  $c$-band detection on 2460251 shows the star approximately 0.3~mag brighter than its typical quiescent value. This slow fading to quiescence from the superoutburst was also seen in the TESS data from 2020 \citep{anousha23}. Overall, the length of the outburst appears typical for SU~UMa type stars \citep{thorstensen02}.

\begin{figure}
    \centering
    \includegraphics[width=\columnwidth]{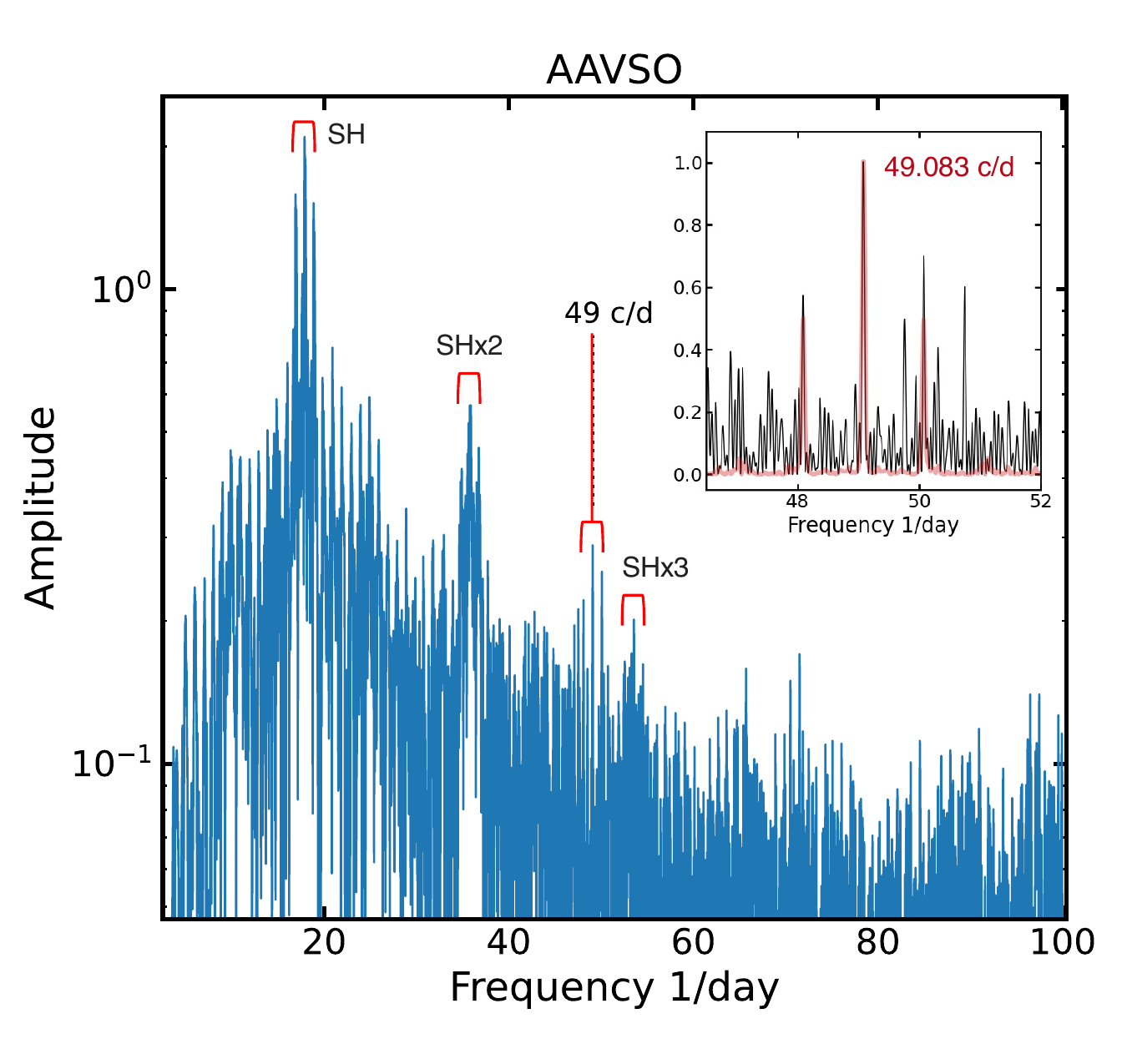}
    \caption{Lomb-Scargle periodogram of the time-resolved photometry from the AAVSO database. Only observations obtained during the bright phase of the outburst were included in the analysis. The superhump modulation and its harmonics are marked on the plot. The 29~minute (49~\cycleday ) periodicity is clearly detected. The inset displays the frequency region around 49~\cycleday . The thick red line is the window function for a 49.083~\cycleday\ sinusoid sampled in the same way as the AAVSO data.}  
    \label{aavso_pow}
\end{figure}

Periodograms of the time-series photometry were calculated using the Lomb-Scargle technique \citep[L-S hereafter;][]{Lomb76, Scargle82}. To avoid small zero-point differences between the many detector/filter combinations in the AAVSO data, we subtracted a second-order polynomial fit from each night of data for each observer. The observations were combined and sorted in time to generate a flattened time-series covering 18 consecutive nights over the brightest phase of the outburst. The resulting L-S periodogram is shown in Figure~\ref{aavso_pow}.  Subtracting the polynomial from each night artificially reduced the power at the lowest frequencies. The daily gaps in the photometry resulted in a window function with two satellite peaks separated by $\pm1$~\cycleday\ from the central frequency and this is displayed in the inset to Figure~\ref{aavso_pow}. The superhump oscillation (SH) and its harmonics are clearly visible in the periodogram. The 49~\cycleday\ periodicity discovered by \citet{anousha23} is also clearly detected. Its amplitude relative to the SH harmonics is similar to that seen in the TESS photometry from the 2020 outburst. The frequency of the signal peak in the AAVSO data is measured to be 49.083$\pm 0.014$~\cycleday , which is consistent with the 49.085$\pm 0.005$~\cycleday\ (29.34 $\pm$ 0.01~min) periodicity found in the TESS data. The stability of the signal implies that it is related to the rotation of the WD in the system.

\begin{figure}
    \centering
    \includegraphics[width=\columnwidth]{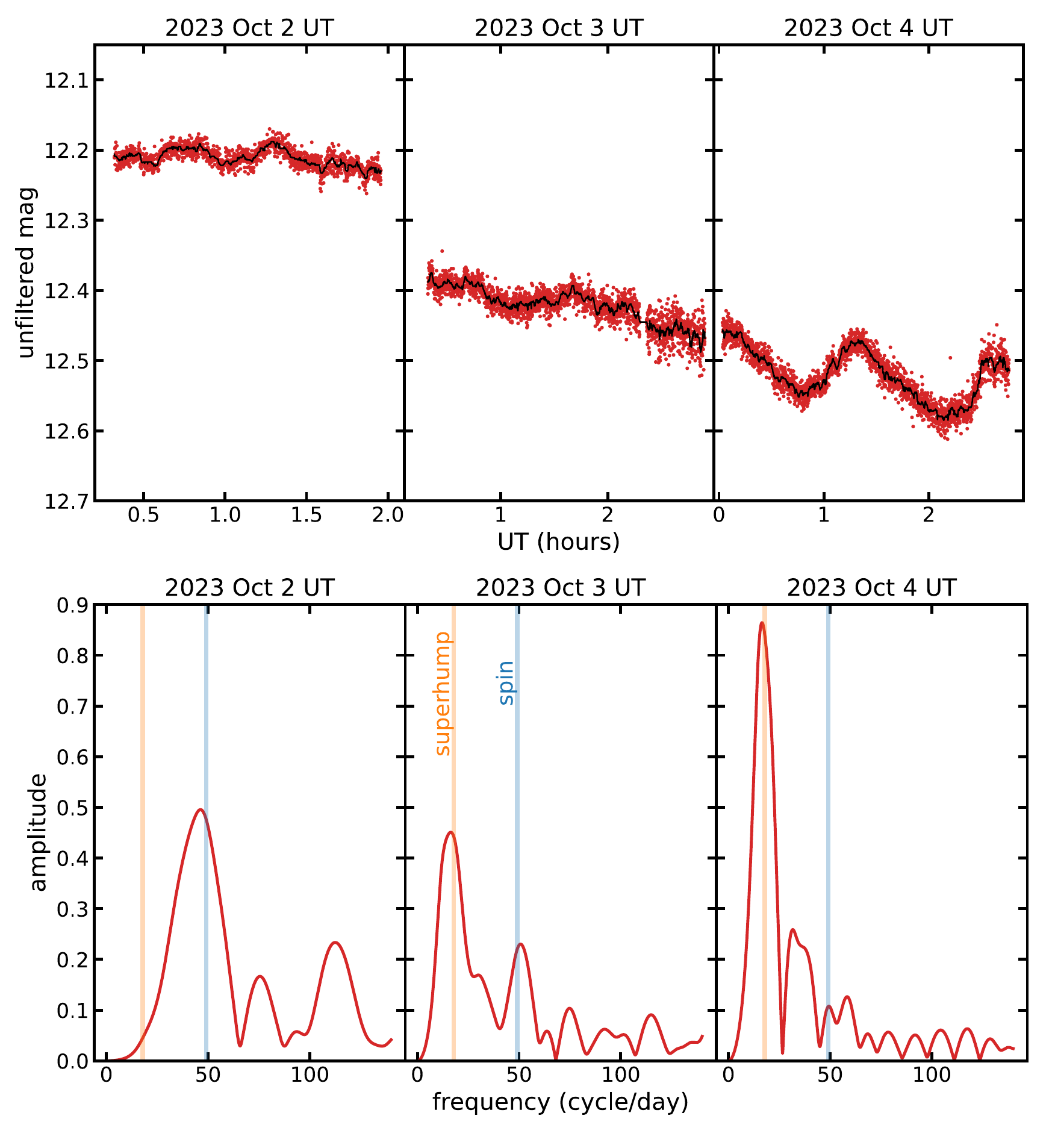}
    \caption{{\bf Top:} Fast cadence photometry of V844~Her during the first three nights of the 2023 superoutburst. The solid line shows the data after median smoothing with a 60~s moving boxcar. {\bf Bottom:} Periodograms for each of the three nights showing the fading amplitude of the 49~\cycleday\ signal and the growth of the superhumps.  }
    \label{kriz_early}
\end{figure}

The SLKT observations began 1~day after the onset of the outburst. The three consecutive nights of high-cadence photometry in Figure~\ref{kriz_early} show a gradual fading of V844~Her from $\sim$12.2~magnitude to $\sim$12.5 magnitude. The L-S periodograms for each night show the decay of the 49~\cycleday\ signal and the development of the SH signal. On the first night of SLKT observation, the 49~\cycleday~signal was the strongest, with little evidence of the SH signal. On the second night, the SH signal is stronger, and by the third night, the 49~\cycleday~signal was no longer detectable.

\begin{figure}
    \centering
    \includegraphics[width=\columnwidth]{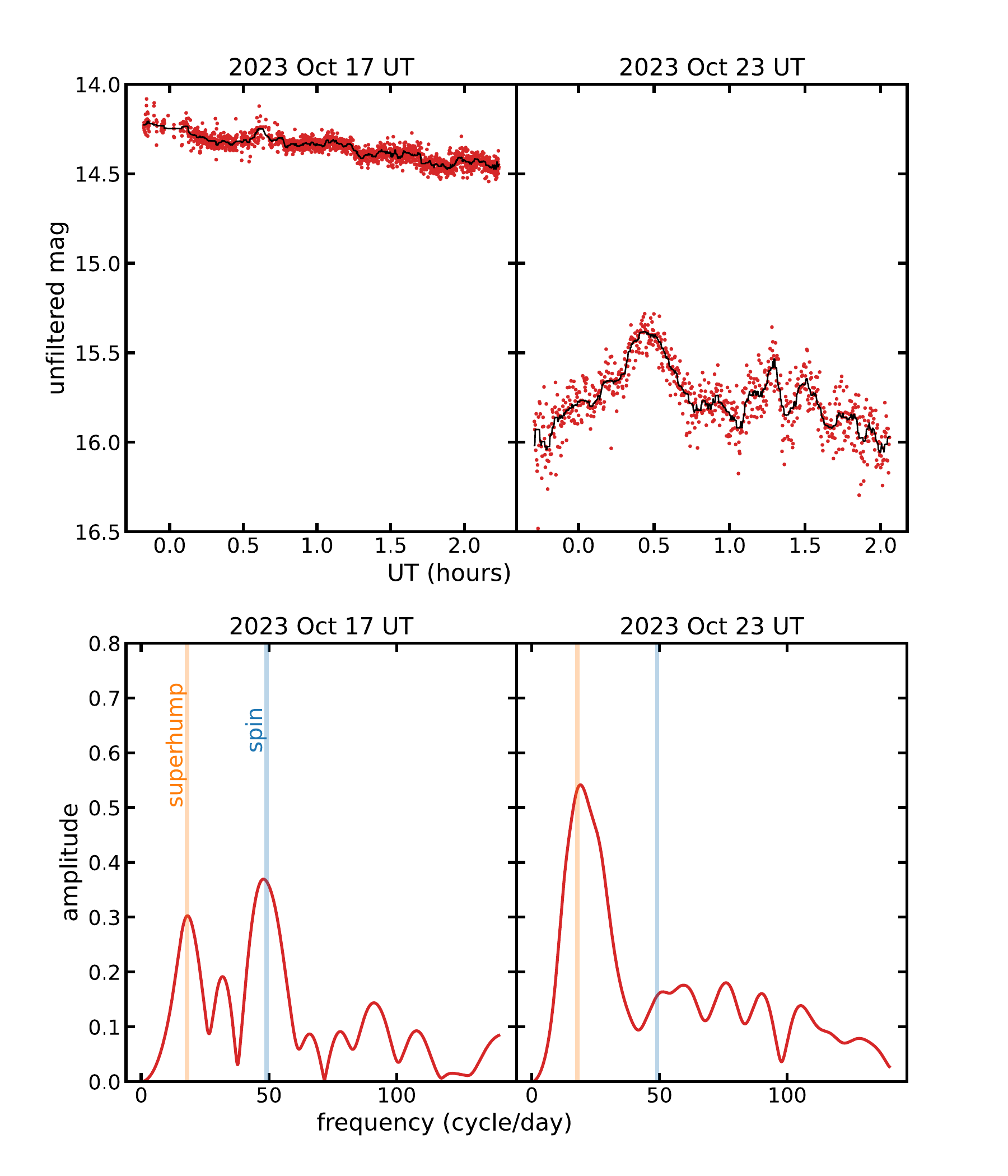}
    \caption{{\bf Top:} Fast cadence photometry of V844~Her during the end of the 2023 superoutburst. The solid line shows the data after median smoothing with a 60~s moving boxcar. {\bf Bottom:} Periodograms for each of the two nights showing the presence of the 49~\cycleday\ signal two weeks after the outburst began. By three weeks after the burst, the 49~\cycleday\ signal is not detected.  }
    \label{kriz_late}
\end{figure}

SLKT photometry was also obtained at the end of the bright phase of the outburst, and during the decline to quiescence. Figure~\ref{kriz_late} displays the light curve and its periodogram at 16 days after the start of the outburst. The 49~\cycleday\ signal is comparable to the SH amplitude at this time. This was also seen in the 2020 TESS photometry immediately prior to the rapid fading at the end of the bright outburst phase. The final SLKT time-series is from 22~days after the start of the outburst during the final fade to quiescence. Here, the periodicity is dominated by the SH oscillations or orbital modulation with little contribution from the 49~\cycleday\ signal. 

\subsection{X-ray Variations}

\begin{figure}
    \centering
    \includegraphics[width=\columnwidth]{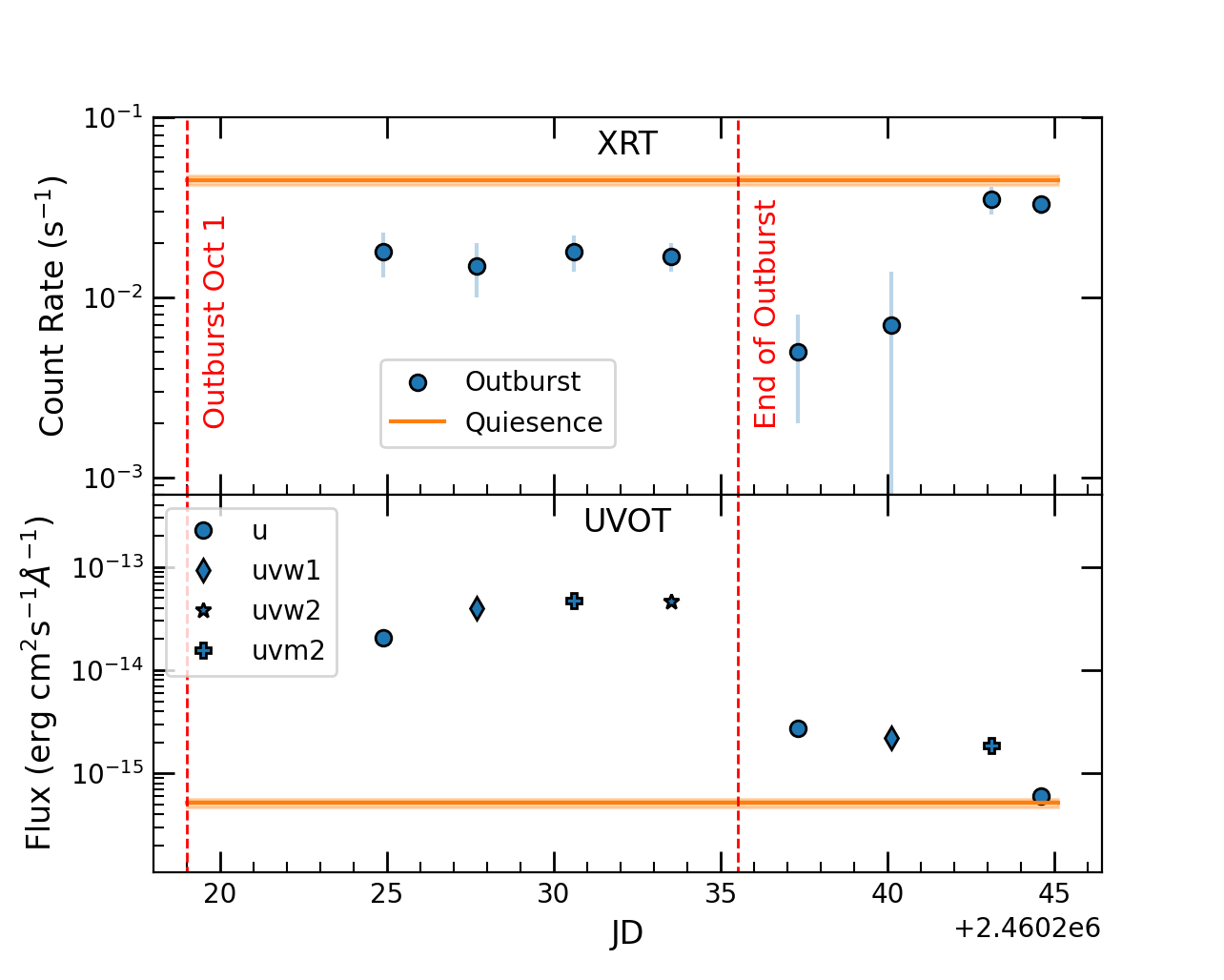}
    \caption{{\bf Top:} The V844~Her X-ray count rate of from XRT over the superoutburst with the quiescent count rate from 2016 (orange band) shown for reference. {\bf The onset and end of the bright optical outburst are indicated by the vertical dashed lines}. {\bf Bottom:} The UVOT flux over the outburst with the quiescent measurement (measured with a u-band filter) from 2016 shown as an orange band. }
    \label{xrtuvot}
\end{figure}

The XRT light curve during the superoutburst is shown in the top panel of Figure \ref{xrtuvot} and compared with the quiescent fluxes. Swift visited V844~Her six days after the onset of the outburst and found the XRT count rate to be half that of the 2016 quiescent level. The XRT count rate remained fairly constant for three more visits covering 15~days after the outburst began. At the end of the optical plateau phase, the X-ray count rate further fades to $\approx$15\%\ of the quiescent level.  The final two Swift visits more than 23 days after the start of the outburst find the XRT count rate has recovered to the quiescent level. Unfortunately, the Swift XRT count rate from V844~Her is typically 0.02~counts~s$^{-1}$ or less, meaning that detailed spectral analysis is not possible. 

The X-ray variations in short-period CVs display a wide range of properties during superoutbursts. WZ~Sge type systems are a sub-class of DN that possess very short orbital periods and that only display superoutbursts and no ``normal" outbursts. \citet{neustroev18} found that the X-ray fluxes of WZ~Sge show a factor of 5 enhancement relative to their quiescent levels . These systems also show changes in their X-ray properties when transitioning from Stage~A to Stage~B superhumps and at the end of a superoutburst. This connection is hard to understand given that the X-rays are thought to originate in the inner disk, while the dynamical instability involves the outer disk. 

In contrast, the prototypical CV SU~UMa displays a factor of 4 to 5 drop in X-ray flux during normal outbursts and the flux recovers as the optical brightness fades \citep{collins10}. Fitting the RXTE data, \citet{collins10} found that the overall X-ray luminosity declines in outburst and the spectrum gets softer. A superoutburst of SU~UMa has not been studied in X-ray wavelengths.

Outbursts have been seen in a large fraction of IPs with orbital periods of less than two hours \citep{hameury17}. Unlike typical SU~UMa CVs, IP outbursts tend to last just a few days or less, so it is not clear if they result from the standard thermal disk instability model (DIM). \citet{littlefield22} suggested that short rapid bursts from V1025~Cen result from magnetically gated accretion. Of the systems with orbital periods less than 2~hours, only IPs V455~And and CC~Scl display classic DIM superoutbursts. V455~And is a high-inclination system with an extremely low X-ray flux \citep{mukai23}, making it less than ideal as a comparison system.  The WD spin modulation in CC~Scl was discovered during a disk superoutburst, suggesting that it is a good system to compare with V844~Her.

\citet{woudt12} analyzed Swift XRT observations of CC~Scl that began in the middle of a superoutburst and continued for several days after fading, but no Swift observations during quiescence have been obtained. The total XRT count rate for CC~Scl showed a modest factor of two decline at the end of the superoutburst. In contrast, V844~Her displayed a significant drop in the X-rays at the end of the optical plateau, with a decline of the XRT count rate by a factor of three to four.  

\citet{woudt12} modelled the CC~Scl X-ray spectra and found a sharp drop in the hydrogen column and covering fraction at the end of the superoutburst. After correcting for the internal absorption of soft X-rays, they found that changes in the bolometric X-ray luminosity of CC~Scl were similar to the optical variations at the end of the superoutburst. That is, the bolometric luminosity faded by a factor of five as the light curve fell off the plateau. 
The hard X-ray flux (2.5-10 keV) was five times stronger during the outburst than after the optical fading. The soft X-rays (0.3-2.5 keV) were suppressed at the start of the observing sequence, but steadily increased over the last half of the outburst. The soft flux continued to increase just as the bright optical outburst ended. \citet{woudt12} suggest that variations in the accretion rate combined with changes in the hydrogen column are important in determining the total X-ray flux in IPs. 

\begin{figure}
    \centering
    \includegraphics[width=\columnwidth]{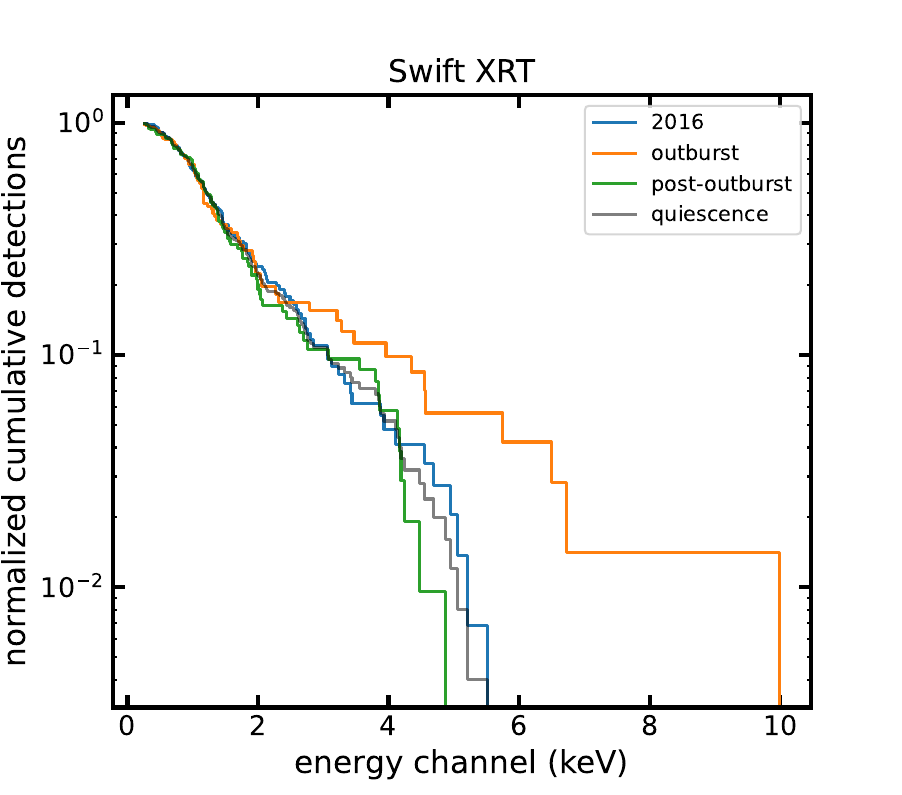}
    \caption{The cumulative energy distribution of detections for XRT observations of V844~Her. The 2016 observations (blue line) were during quiescence, and the post-outburst observations (green) are the sum of the final two Swift visits. The combination of 2016 and the final two visits are displayed as a grey line labeled ``quiescence''. The combination of the four Swift visits during the bright optical phase (orange line) is labeled ``outburst''. During the optical outburst, the fraction of detections with energies $>3$~keV appear larger than during quiescence.       }
    \label{cumulative}
\end{figure}

The very low XRT count rate during the outburst of V844~Her makes a detailed spectral analysis difficult. However, we can construct a  
cumulative distribution of detected photon energies using the pulse invariant (PI) parameter from the XRT event files. We compare the cumulative energy distributions (Figure~\ref{cumulative}) during the bright outburst phase with the 2016 quiescent observation and the final two Swift visits after the outburst. There appears to be a larger fraction of detections with energies $>3$~keV during the outburst phase than during quiescence. A Kolmogorov-Smirnov test gives a p-value of 0.2, which supports, but does not definitively indicate, that the outburst and quiescent detections are drawn from different distributions.

\subsection{UV Light Curve}

The UVOT light curve during the superoutburst is displayed in the lower panel of Figure \ref{xrtuvot}. The first UVOT Swift measurement shows the UV flux to be forty times the quiescent level. Caution is needed in interpreting the light curve as UVOT filters varied over the visits. We note that UVOT grism spectra of GW~Lib taken during the outburst showed a fairly flat continuum between 220 and 320~nm in $F_\lambda$ \citep{byckling09}. We conclude that the UV flux from V844~Her was roughly steady as the outburst progressed. Around the end of the bright optical plateau the UV flux drops by a factor of 10.  The flux then remains roughly steady at this fainter state for a week before the final Swift visit finds that it has returned to 2016 quiescent level. 

The V844~Her UVOT light curve is similar to that seen from CC~Scl at the end of its superoutburst observed by Swift. The final Swift visit to V844~Her suggests a return to quiescence 9~days after the end of the superourburst. This is not seen in CC~Scl, but Swift observtions of CC~Scl ended just 5~days after the optical fading of its superoutburst.

\begin{figure}[b]
    \centering
    \includegraphics[width=\columnwidth]{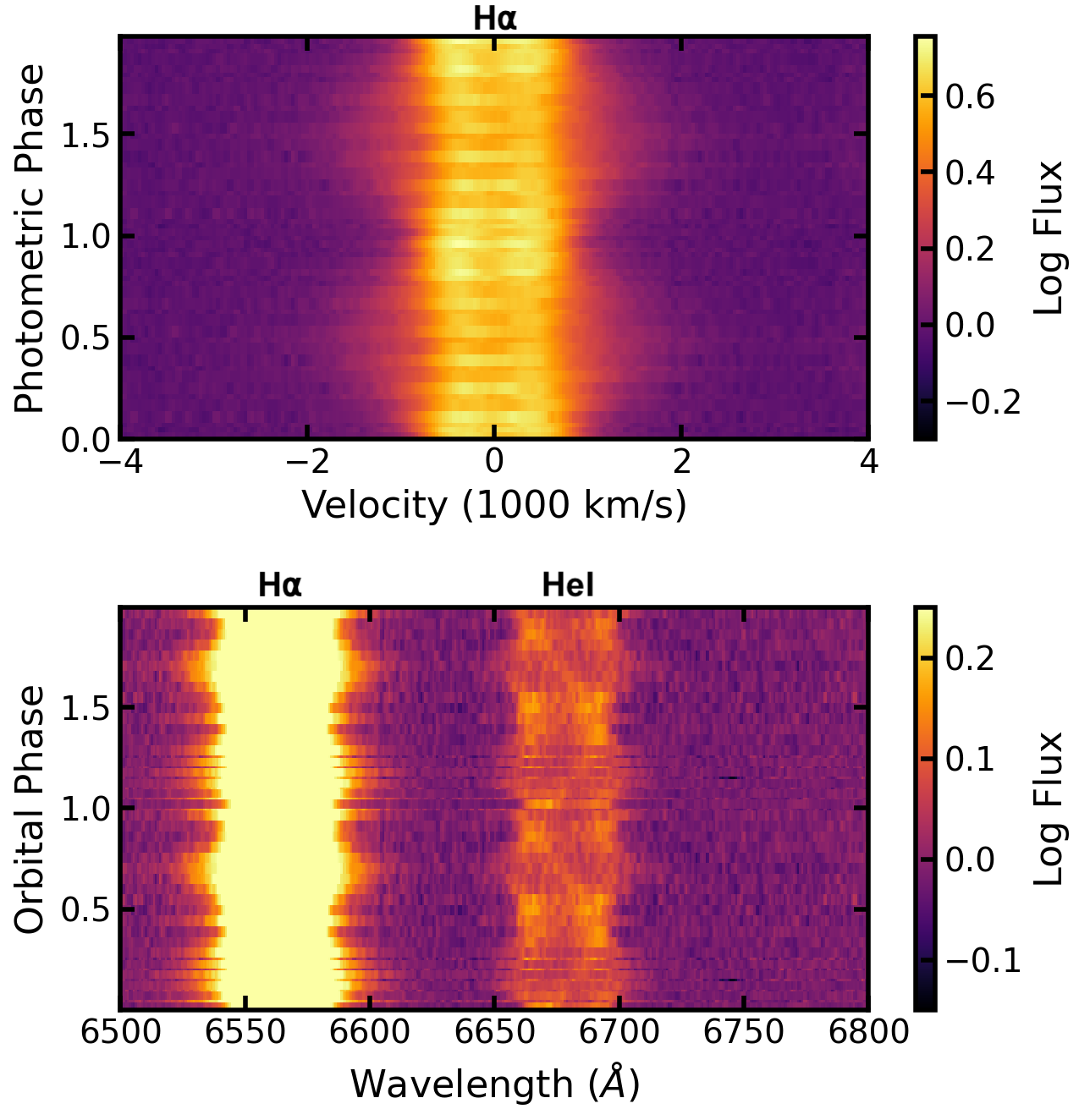}
    \caption{{\bf Top:} The spectral time-series at H$\alpha$ is phased on the TESS photometric period (29.34~min). High-velocity line wings are seen extending out 2000~\kms\ from the rest wavelength.  The bright central emission displays a double peak coming from the red and blue-shifted sides of an accretion disk. {\bf Bottom:} The spectral time-series at H$\alpha$ and He~I is phased on the orbital period (78.69~min). The H$\alpha$ emission core is over-exposed to better show the variation in the wings. A weak emission spot is seen making an S-curve within the He I line over an orbit. } \label{myspec}
\end{figure}

\section{Analysis of Quiescence}

\subsection{Emission Line Variability in Quiescence}

To view changes in the optical emission lines, we normalized each of the LBT spectra by their continuum level. This was accomplished by fitting a polynomial to the continuum and then dividing the result into each spectrum. We phased the spectra on the orbital period, 78.69~min, and display the stacked spectra in the bottom panel of Figure \ref{myspec}.  Variable, high-velocity wings on the H$\alpha$ and He~I emission lines are clearly seen, however, variations in the wings are not in phase with the orbit. An S-wave in the He~I line is seen matching the orbital period.

\begin{figure}
    \centering
    \includegraphics[width=\columnwidth]{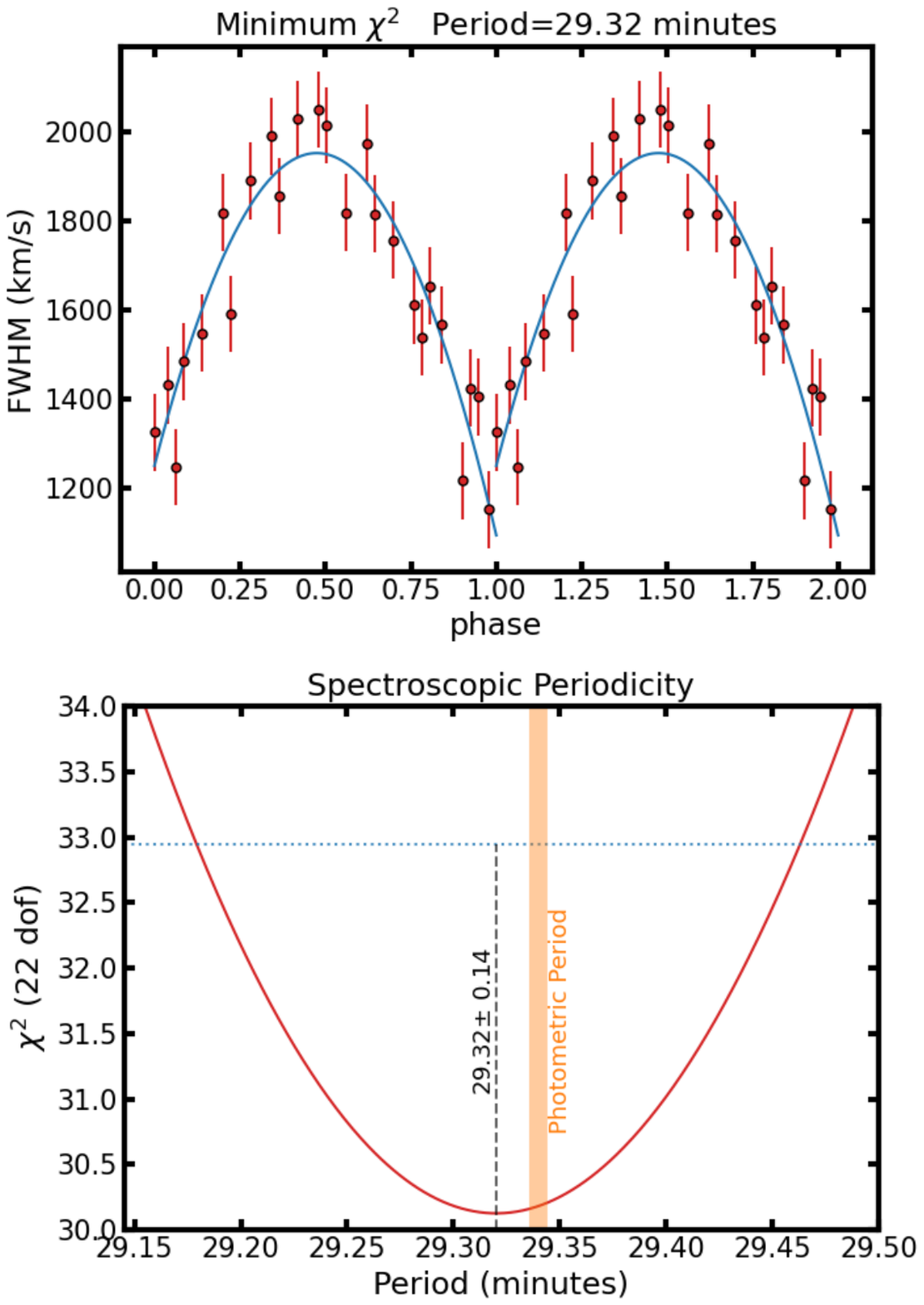}
    \caption{{\bf Top:} The width of the H$\alpha$ emission line in quiescence as a function of phase for the period giving the lowest $\chi^2$ value. The solid blue line displays the best fit quadratic model.  {\bf Bottom:} The change in the $\chi^2$ parameter as the spectroscopic period is varied. The lowest $\chi^2$ parameter was found for a period of 29.32$\pm$0.14 minutes and is marked by the dashed line. The dotted line displays the 1-$\sigma$ uncertainty from the $\chi^2$ minimum. The photometric period measured during superoutbursts is shown as a vertical band.} \label{spec_period}
\end{figure}

To constrain the periodicity in the high-velocity features, we fit the wings of the H$\alpha$ emission in each spectrum with a Gaussian that excluded the bright inner core of the line. The resulting full-width at half-maximum (FWHM) of the fits varied between 1200 and 2000~\kms\ with an uncertainty of 90~\kms\ on each measurement. We fit a quadratic function to the FWHM versus phase for a range of possible periods and minimized the $\chi^2$ parameter between the data and the quadratic model as shown in Figure~\ref{spec_period}. The best fit was found for a period of 29.32$\pm 0.14$~minutes with a minimum $\chi^2$ parameter of 30 for 22 degrees of freedom. This spectroscopic period is consistent with the TESS photometric period of 29.34$\pm 0.01$~minutes measured by \citet{anousha23} during the 2020 superoutburst, and the period found in the AAVSO data during the superoutburst of 2023.

While TESS and ground-based observations could not detect the 29-minute periodicity when V844~Her was in quiescence, the wings of spectral lines in quiescence display the same periodicity as discovered during superoutbursts. The spectral time-series for H$\alpha$, phased on the photometric period is shown in the top panel of Figure~\ref{myspec}.  The cycling of the high velocity line wings is difficult to detect from photometric measurements alone, because the total flux of the Balmer emission is nearly constant (see the lower panel of Figure~\ref{ave_spec}). This is consistent with the source of the broad emission being Doppler shifted to lower velocities, so contributing to the total line flux throughout the 29-minute cycle. 

The high-velocity emission wings are seen extending to $\pm 2000$~\kms , and are likely the result of accretion onto a magnetic, asynchronously spinning WD. We conclude that V844~Her is an IP. 

It is very likely that the spin period of the WD corresponds to the 29-minute spectroscopic modulation. A few IPs show a dominant periodicity at double the spin frequency, meaning the true spin period may be twice as long as the observed modulation.   Given the limited data for this system, we cannot rule out a WD spin period of 58-minutes. See further discussion of the WD spin period in Section \ref{spin}.

\subsection{Orbital Diagnostics}

We infer the WD's orbital motion from the shifts in wavelength of the emission line wings. We applied the “double Gaussian” method developed by \citet{shafter83} and \citet{schneider80}, which consists of convolving the emission line of interest with positive and negative Gaussian functions separated by a fixed wavelength, $S$. The velocity at which the convolution function goes to zero is an estimate of the centroid of the emission from the inner disk, approximating the motion of the WD. We used a Gaussian width of $\sigma=10$ \AA. To determine the best Gaussian separation, we calculated the velocity curve using full separations ranging between $55<S<90$ \AA. A sinusoid is fit to the velocity curve from each value of $S$, and the fit quality is determined by their $\chi^2$ parameter. The parameters of the sinusoid fit were the amplitude, offset, and phase. The orbital period was fixed at the known value of 1.31~hr due to limited phase coverage of the spectroscopy.

One concern with applying the double-Gaussian method on this system is that the highest velocity emission varies on the 29~minute photometric period that we associate with magnetic accretion. As this periodicity is significantly shorter than the orbit, its variations could compete with the velocity amplitude variations caused by the orbital motion of the WD. However,  emission varying on the 29~minute period appears symmetric on the red and blue sides of the H$\alpha$ line, thus it is roughly cancelled out in the double-Gaussian convolution.

\begin{figure}
    \centering
    \includegraphics[width=\columnwidth]{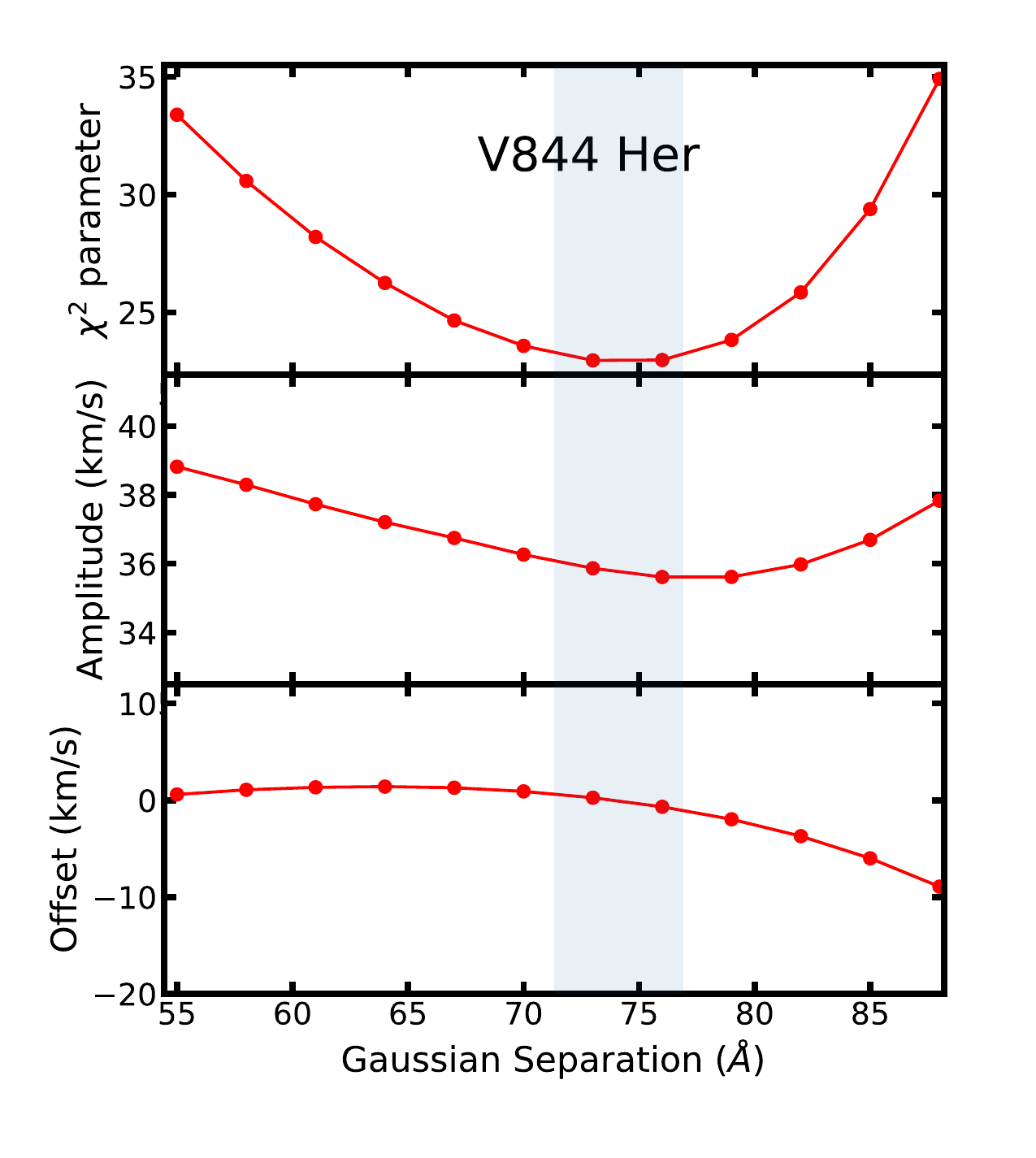}
    \caption{The results of the double-Gaussian method applied to V844~Her spectra. The top panel shows the varying $\chi^2$ parameter of the sinusoidal fit to the velocity curve as a function of the full Gaussian separation. The lowest $\chi^2$ occurs at a separation of 74~\AA .  The middle panel displays the velocity amplitude of each fit. The lower panel shows the offset of the sinusoid which corresponds to the systemic velocity of the binary center of mass.
    \label{diagnostic}}
\end{figure}

The results of the double-Gaussian method applied to the H$\alpha$ emission line are shown in Figure~\ref{diagnostic}. The $\chi^2$ parameter reached a minimum at $S=74$~\AA\ where the velocity amplitude was 36$\pm 2$~\kms . The velocity amplitude and offset velocity are relatively constant over the whole range of Gaussian separations.

\begin{figure}
    \centering
    \includegraphics[width=\columnwidth]{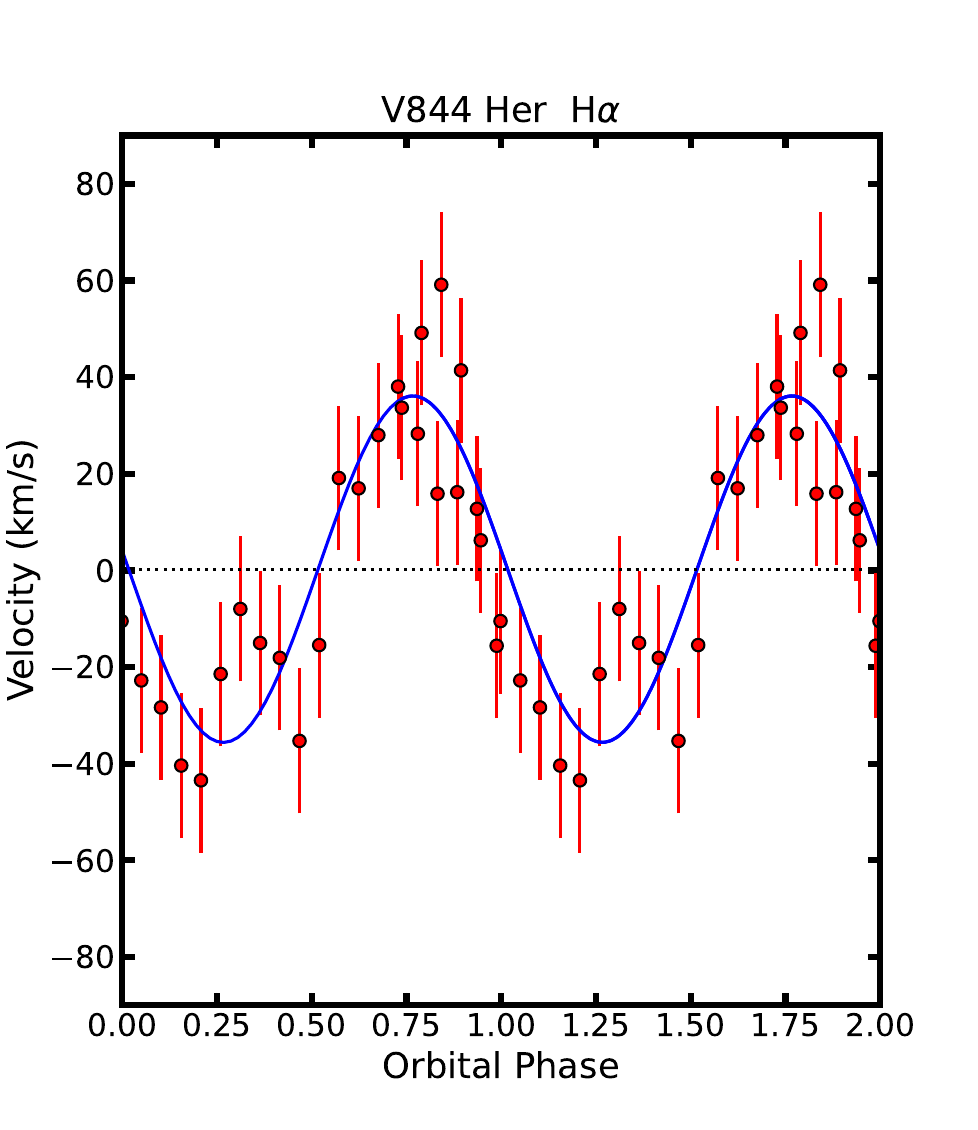}
    \caption{The center of the double-Gaussian applied to the H$\alpha$ line as a function of orbital phase for the fit with the lowest $\chi^2$ parameter. 
    \label{amplitude}}
\end{figure}

The best fit orbital velocity curve is shown in Figure~\ref{amplitude}. The time of zero orbital phase (inferior conjunction of the secondary star) is estimated to be MJD=60112.353$\pm 0.004$. 

\subsection{Tomography}

\begin{figure*}
    \centering
    \includegraphics[width=1.3\columnwidth]{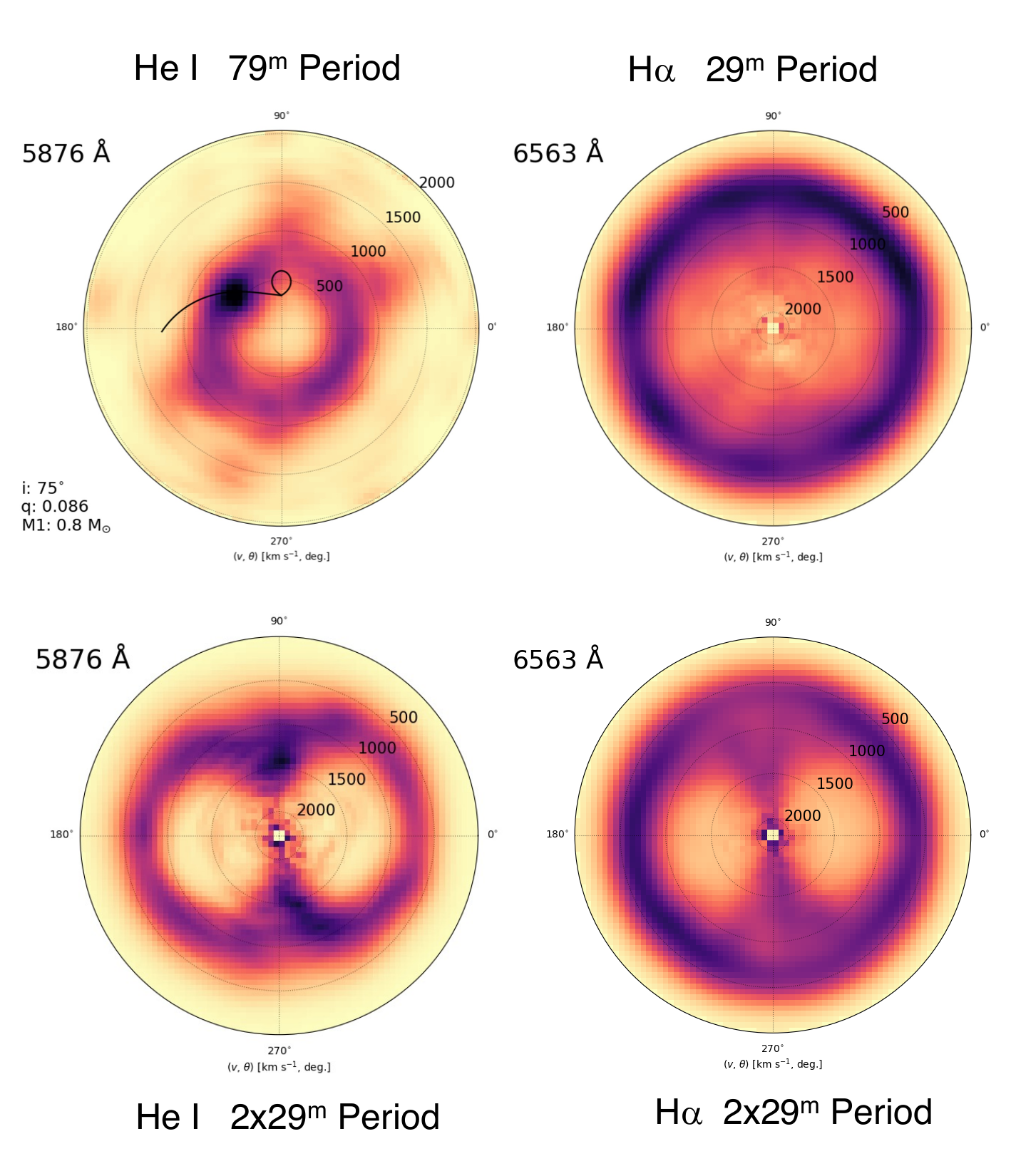}
    \caption{Tomograms constructed from the LBT spectral time-series obtained during quiescence. {\bf Top Left:} Standard tomogram of the He~I 587.6~nm emission line phased on the 78.7 minute binary orbital period. The S-wave feature seen in the spectral time-series is consistent with a hotspot where the stream strikes the disk. {\bf Top Right:} Inside-out projection from the H$\alpha$ emission line phased on the 29 minute photometric period. {\bf Bottom:} Inside-out projections of the He~I and H$\alpha$ emission lines phased on twice the photometric period. 
    \label{tomography}}
\end{figure*}

We constructed tomograms \citep{marsh88} from the LBT time-series spectroscopy using the "doptomog" code provided by \citet{kotze_phd}. The He~I emission features displayed a weak S-wave component either from a hot spot on the accretion disk or a heated face of the secondary star.  The zero point of the orbital phase was set to the value derived from the double-Gaussian analysis. The overlay of the secondary and stream required estimates of the WD mass, binary mass ratio and the inclination. From the characteristics of its superhumps, \citet{kato22} estimated the binary mass ratio in V844~Her to be $q=0.086\pm 0.002$ and the WD mass was assumed to be 0.8~M$_\odot$. The resulting tomogram is shown in Figure~\ref{tomography}.  The S-wave produces a bright spot at 600~\kms\ and $\theta\approx 135^o$, and likely corresponds to where the stream from the secondary strikes the accretion disk.

An inside-out projection \citep{kotze16} was constructed from the H$\alpha$ emission feature. The initial tomogram was phased on the 29-minute photometric period (upper-right of Figure~\ref{tomography}), but no features inside the accretion disk are seen. We also constructed inside-out tomograms assuming the WD spins at twice the photometric period, as shown in the lower panel of Figure~\ref{tomography}.

\section{Discussion}

\subsection{A New Intermediate Polar}

The spectra obtained during quiescence show evidence of high-velocity wings modulated at the 29~min period that is detected in photometry only during superoutbursts. This confirms the presence of magnetic accretion and that V844~Her is an IP.  

As noted in \citet{anousha23}, other indicators of magnetic accretion are not present in the quiescent state of V844~Her. There is no detected modulation during quiescence at the 29-min period by TESS or extensive ground-based photometry. XMM observations did not show any X-ray modulation, while strong signals are seen in known IPs with comparable orbital periods. Our optical spectroscopy of V844~Her shows only very weak He~II emission in quiescence.

The lack of  a 29-min optical continuum variation and X-ray modulation in V844~Her during quiescence may suggest that the accreting gas is not generating a high-temperature shock near the surface of the WD. This could be due to a very low specific accretion rate that results in a collisionless flow known as the bombardment regime \citep{busschaert15}. This is an unlikely proposition for an IP, so the discussion of the bombardment regime is considered in Appendix~A.

Another superoutbursting system with a short orbital period is CC~Scl. The spinning WD in the IP CC~Scl, was identified in photometric variations detectable only during superoutbursts  \citep{woudt12}. Later, \citet{szkody17} detected a clear spin modulation during quiescence in FUV observations from Hubble Space Telescope (HST). 
CC~Scl shows a strong He~II emission line during superoutburst \citep{woudt12} while during quiescence its He~II line is very weak \citep{pena15}.

\subsection{Spin Period} \label{spin}

In the canonical model of IPs, the high-velocity emission is produced by the accretion curtains, which are structures formed by magnetically channeled gas flowing from the disk along the WD's magnetic field lines. Because the accretion curtains corotate with the WD, their emission is a marker of the WD spin period. 

\begin{figure}
    \centering
    \includegraphics[width=\columnwidth]{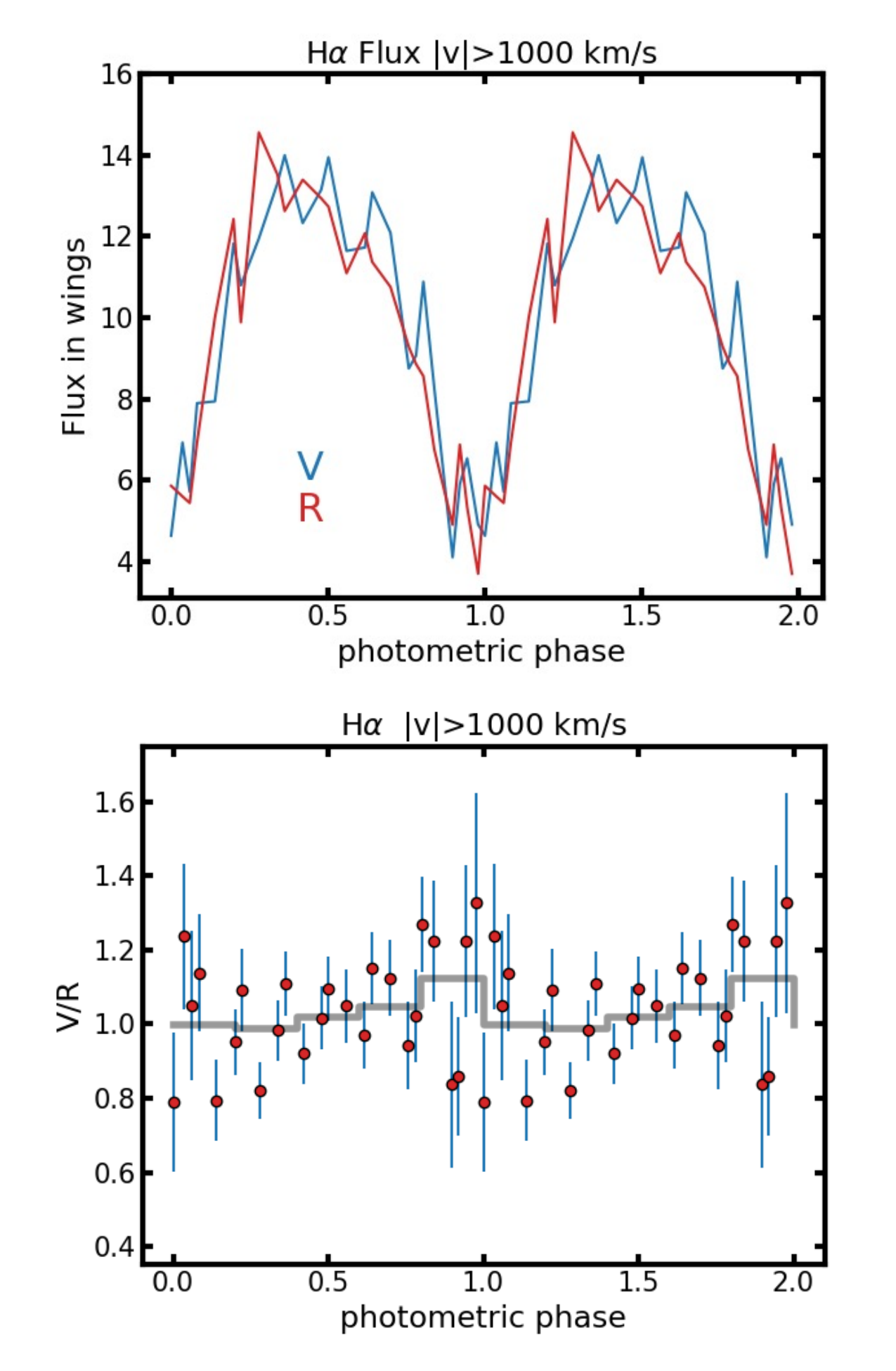}
    \caption{{\bf Top:} The relative flux in the H$\alpha$ emission line wings. The blue-shifted wing ($V$) was measured for velocities less than $-1000$~\kms , and the red-shifted wing ($R$) measured for velocities greater than 1000~\kms . A cycle is repeated for clarity. {\bf Bottom:} The $V/R$ ratio phased on the 29.3 minute photometric period . The average of the $V/R$ ratio is close to unity. The thick gray line shows the weighted mean of the $V/R$ measurements binned by 0.2 in phase. Typically there are 4 to 6 spectra in each bin. The noise in these measurements makes it difficult to tightly constrain the amplitude of any $V/R$ variation with phase.
    \label{vr}}
\end{figure}

A few IPs show a dominant photometric period on the harmonic of their spin period. For example, the optical variations of V405~Aur are primarily seen at twice its WD spin frequency \citep{still98}. \citet{evans04} suggest that a double-peaked spin pulse results from a geometry where the dipole inclination is sufficiently high that both curtains are nearly equally visible.
One of the most significant implications of this scenario is that it can introduce a factor-of-2 ambiguity in the spin period that we call the Evans and Hellier ambiguity.

EX~Hya, like most IPs, generates a single optical pulse per spin cycle. It also displays a spectroscopic asymmetry over a spin cycle. The section of the curtains nearest the disk dominates the accretion flow (see Figure~11 in \citet{hellier87}), so over a spin period, the flow along the bright portion of the curtain is viewed at different angles due the inclined dipole axis. Thus, the line wings are not viewed symmetrically. This is seen as a variation in the violet-to-red flux ratio ($V/R$) of the emission lines over a spin cycle. \citet{hellier87} found an amplitude of 0.10 in the $V/R$ ratio over a spin cycle.

To test the degree of asymmetry in the emission wings of V844~Her, we measured a $V/R$ ratio of the H$\alpha$ line phased on the photometric period. The average $V/R$ ratio for V844~Her in Figure~\ref{vr} is close to unity, but the noise in the measurements makes it difficult to constrain the amplitude of any variation with phase. Randomizing the phases and fitting a sinusoid results in an amplitude of 0.1 or greater in 40\%\ of the trials.  With the current data, we cannot constrain the degree of symmetry of the line wings in V844~Her with sufficient accuracy to resolve the Evans and Hellier ambiguity.

Here, we adopt a WD spin period for V844~Her of 29.3~minutes, although further observations could revise this value.

\subsection{X-Rays in Quiescence}

From the serendipitous XMM observation of V844~Her, \citet{anousha23} estimated an X-ray luminosity of 1.1$\times 10^{31}$~\ergsec. We can compare this with the soft X-ray ($<10$~keV) luminosities from established IPs compiled by \citet{mukai23}. Generally, V844~Her has a factor of 3 to 5 times lower soft X-ray luminosity than IPs with similar orbital periods. One exception is CTCV~J2056-3014 which has a luminosity comparable to V844~Her. As an IP, V844~Her would fall in the subclass of low-luminosity IPs \citep[LLIP;][]{pretorius14}
\footnote{We note that V455~And stands out among the short-period IPs with an exceptionally low soft X-ray luminosity of only 5.5$\times 10^{28}$~\ergsec , possibly due to its highly inclined orbit.}. 

With an absolute magnitude of $M_G=10.0$ \citep{canbay23}, V844~Her is typical for the shortest period IPs. As done by \citet{mukai23}, we construct a disk brightness index (DBI) of $-0.184$ for V844~Her, which is typical for short-period IPs.

\subsection{X-Rays in Outburst}

Many dwarf novae systems show a decrease in X-ray flux during outburst. This is thought to be due to the boundary layer becoming optically thick. \citet{fertig11} show an anti-correlation between 2-10~keV X-ray flux and optical brightness in outbursting dwarf novae. But a few short-period CVs, such as WZ~Sge, show the opposite behavior: the X-ray flux is enhanced during superoutbursts \citep[e.g.][]{neustroev18,collins10} . The diversity in X-ray properties during disk outbursts in non-magnetic CVs is quite confusing. 

CC~Scl is one of the few IPs that experience long superoutbursts like those found in SU~UMa systems, thus CC~Scl provides a context to understand the V844~Her Swift XRT light curve. After correcting for the hydrogen column density, \citet{woudt12} found significant bolometric X-ray production during the CC~Scl superoutburst. This likely resulted from the increase in accretion onto the WD. However, they also found during the disk outburst, a large absorbing hydrogen column that preferentially removed soft X-rays from the spectrum. The significant increase in hard X-ray flux during the superoutburst is masked by the loss of soft X-rays caused by the larger hydrogen column.  After the outburst, this spectral ``see-saw'' reverses, and the soft X-rays escape while the harder accretion-generated X-rays fade. This naturally leads to a rather modest overall change in the total detected X-ray counts during a superoutburst in IPs.

The spectral see-saw has been seen at work in FO~Aqr. While not a dwarf nova, in 2016 FO~Aqr faded by more than two magnitudes from its typical bright-state \citep{littlefield16}. \citet{kennedy17} found only a small change in the total count rate in Swift XRT data taken before and during the low-state. But their Figure~7 confirms that during the low-accretion state, the XRT spectrum peaks at low energies, while in its bright state the peak shifts to harder X-rays.

The X-ray properties of V844~Her over the superoutburst are roughly the same as seen for CC~Scl \citep{woudt12}. We suspect that as V844~Her enters superoutburst, the enhanced accretion rate onto the WD generates more X-rays than during quiescence. But the rise in the absorbing column \citep{knigge97,baskill01} from the disk outburst removes flux in soft X-rays resulting in a slight decrease in the overall XRT count rate relative to the quiescent rate (see Figure \ref{xrtuvot}). Then, as seen in CC~Scl, the end of the bright optical outburst means a lower accretion rate and a sharp decline in hard X-rays. This leads to a further fading of the XRT count rate. Finally, a week or two after the end of the superoutburst, the hydrogen column from the disk outburst has sufficiently dissipated to allow soft X-rays to escape during quiescence. At this point, the XRT count rate has returned to the quiescent level measured in 2016. This time-scale is supported by the TESS detection of the 49~\cycleday\ signal continuing for a week after the outburst ended before finally fading away \citep{anousha23}.


\subsection{Evolution}

\begin{figure}
    \centering
    \includegraphics[width=\columnwidth]{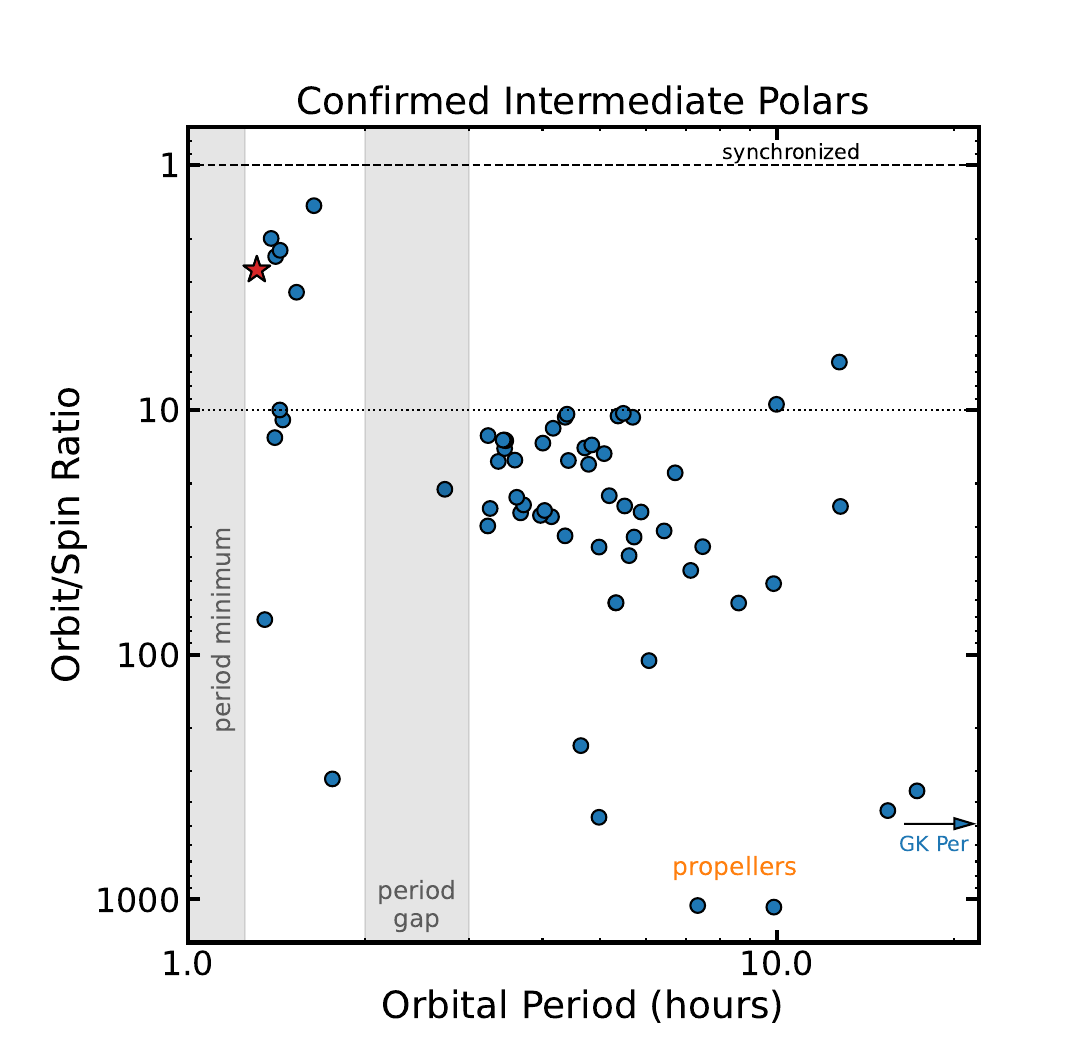}
    \caption{The orbit to spin period ratio versus binary orbital period of confirmed IPs. The star represents the location of V844~Her. The vertical band between 2 and 3 hours shows the CV period gap and the band at the far left indicates the 75~min period minimum for hydrogen-rich CVs. The confirmed propellers AE~Aqr and J0240 are marked. GK~Per has an orbital period of 2.0~days.
    \label{mukai}}
\end{figure}

With its very short orbital period and a long WD spin period, V844~Her is a rare IP that is not far from synchronization (see Figure~\ref{mukai}).  With a WD spin period of 29.3$\pm0.1$~minutes, the orbital to spin period ratio for V844~Her is 2.68$\pm 0.01$.  Only five other confirmed IPs have orbit/spin ratios less than 5.

The distance from the white dwarf where the angular velocity of its magnetic field matches the Keplerian orbital speed is called the co-rotation radius, $R_{co}$ \citep[e.g.][]{campbell11}. Within $R_{co}$, the disk gas is moving faster than the field, which tends to add angular momentum to the WD. The orbit to spin ratio in V844~Her implies a co-rotation radius that is 52\%\ of the binary separation, meaning the majority of the disk is well within $R_{co}$. Thus, we can safely assume that the WD is gaining angular momentum from the disk and spinning up.  Currently, there is no measured period derivative for V844~Her, but EX~Hya, a confirmed IP with a low $P_{orb}/P_{spin}$ ratio, has a well-determined negative $\dot P$ that demonstrates its WD is spinning up \citep{mauche09,beuermann24}.

The large value of $R_{co}$ implies that the system is far from equilibrium. As the WD spins up, the orbit to spin ratio increases and the co-rotation radius moves inward. In V844~Her, this should continue until $R_{co}$ reaches a radius where the magnetic interaction with the disk begins to shed WD angular momentum. \citet{king91} suggested spin equilibrium in IPs with disks occurs for orbit to spin ratios around 10. Indeed, most IPs have $10<P_{orb}/P_{spin}<30$ as seen in Figure~\ref{mukai}. 



\section{Conclusions}

We have obtained new optical photometry during a superoutburst of the SU~UMa type CV V844~Her, as well as X-ray and UV observations during and immediately after the outburt. Combined with time-series spectroscopy in quiescence, we find the following:

\begin{itemize}

\item The 29-minute (49~\cycleday ) periodicity seen in TESS data during the 2020 superoutburst of V844~Her is confirmed in extensive ground-base optical photometry during the 2023 superoutburst. The periodicity is easily detected in the first few days of the outburst, before the superhumps are fully developed. 

\item Time-series spectra of V844~Her obtained during quiescence reveals high-velocity wings on the emission lines modulated on a 29-minute period, thus V844~Her is a new IP as suggested by \citet{anousha23}. With a period of just 1.31~hr, V844~Her has the shortest orbit of any confirmed IP \footnote{https://asd.gsfc.nasa.gov/Koji.Mukai/iphome/catalog/alpha.html}.  

\item The symmetric emission wings in the quiescent spectra suggest the possibility that both accretion curtains are equally visible, and that the spin period of the WD is 58.6~minutes, twice the photometric period. But the current data is unable to constrain the high degree of symmetry needed to resolve the Evans and Hellier ambiguity \citep{evans04}.

\item Swift observations show that the XRT count rate is suppressed by a factor of two during the optically bright phase of the superoutburst when compared with the quiescent phase. This behavior is similar to that observed from the superhumping IP, CC~Scl. The decline in the XRT count rate during a superoutburst is a combination of an increase in hard X-rays from the sudden rise in accretion on to the WD  combined with suppression of soft X-rays from an increased hydrogen column and covering fraction. 

\item The Swift XRT X-ray flux from V844~Her is nearly undetectable after the end of the bright plateau phase, but recovers to the quiescent rate a week later. Likely the hard X-rays drop rapidly when the mass transfer slows, but then it takes several days for the hydrogen column from the disk outburst to clear off and allow the soft X-rays to escape.

\end{itemize}

Photometry during outburst detected a new periodicity that we now show is from an asynchronously rotating magnetic WD in V844~Her. V844~Her is a new member of a sub-class of IPs with very short orbital periods, low X-ray luminosities, and orbit to spin ratios $\ll 10$ (see Figure~\ref{mukai}. There are remaining questions about the IP interpretation for V844~Her. The lack of photometric modulation in the optical and X-rays during quiescence is puzzling. 
Also, the unexplained variation in the amplitude of the spin signal during superoutbursts is difficult to reconcile with accretion from the inner disk.

While possible to detect the photometric periodicity from ground-based observations during outbursts, the relatively long WD spin period and huge superhump oscillations make this difficult. There are likely other IPs hidden in apparently ordinary CVs awaiting discovery.

\bigskip

\begin{acknowledgements}
We acknowledge with thanks the variable star observations from the AAVSO International Database contributed by observers worldwide and used in this research. We thank E. Waagen for organizing the AAVSO alert program.

We acknowledge that this research was funded by a College of Science Summer Undergraduate Research Fellowship (COS-SURF) from the University of Notre Dame. This research was partly funded through NASA grant 80NSSC22K0183.

This work has made use of data from the Asteroid Terrestrial-impact Last Alert System (ATLAS) project. 

We acknowledge the use of public data from the Swift data archive, and we thank the Neil Gehrels Swift observatory for quickly responding to the target of opportunity request for observations of V844~Her in outburst.

We thank K. Mukai and NASA for the comprehensive information on intermediate polars made available to the community.

\facilities{AAVSO, TESS, Swift, LBT}

\end{acknowledgements}

\begin{appendix}

\section{Bombardment Regime}

During an early stage of this work, we were intrigued by the possibility that during quiescence, V844~Her might fall into the `bombardment regime' \citep{kuijpers82} or `collisionless flow regime' \citep{busschaert15}, wherein kinetic energy of the accreting gas is radiated away without a shock and at a significantly lower temperature than seen in high mass-transfer rate magnetic systems. The low temperature accretion on the the WD could explain the lack of optical continuum and X-ray modulation during quiescence \citep{anousha23}. 

Bombardment-regime accretion has been reported previously in polars \citep[e.g., in the so-called low-accretion rate polars; ][]{szkody04} and occurs when the specific mass accretion rate falls below a critical threshold. Although our calculations showed that V844~Her is not in the bombardment regime, we think that the possibility of an IP in the bombardment regime is sufficiently novel that it merits several comments. 

The size of the accreting footprint on the WD determines the ion collision rate for a given total mass accretion rate, and for a given magnetic field strength and mass-accretion rate, this factor would make it easier for disk-accreting IPs to enter the bombardment regime in comparison to polars. The fraction of the WD surface accreting gas in polars is expected to be less than 0.1\%\  \citep{imamura84}. The lower fields and advection from the disk in IPs suggests that the accreting region may be significantly larger, and we place a limit on the footprint at 1\%\ of the WD surface.

On the other hand, the cyclotron cooling time decreases steeply with increasing field strength \citep{busschaert15}, which is problematic for invoking bombardment-regime accretion at the low field strengths typically assumed in IPs. However, the WD in the highly asynchronous system J1344 \citep{littlefield23}, possesses a magnetic field strength of 56~MG. Its orbital period of 1.9~h, and orbit-to-spin ratio of 1.12, are not dissimilar to the spin and orbital properties of V844~Her and serves as a cautionary tale against assuming that IPs are all low-field systems.

Despite the well-established application of bombardment accretion to polars, we are unaware of any papers that have contemplated bombardment accretion in IPs. If these IPs exist, there are at least several interesting implications. In particular, they would be X-ray faint and might lack high-excitation lines in their optical spectra, both of which are normally considered important observational criteria for classifying systems as IPs. Second, the occurrence of outbursts might cause the system to briefly enter a shock regime, so its spin modulation might become detectable during outbursts. 

\begin{figure}[h!]
    \centering
    \includegraphics[scale=0.7]{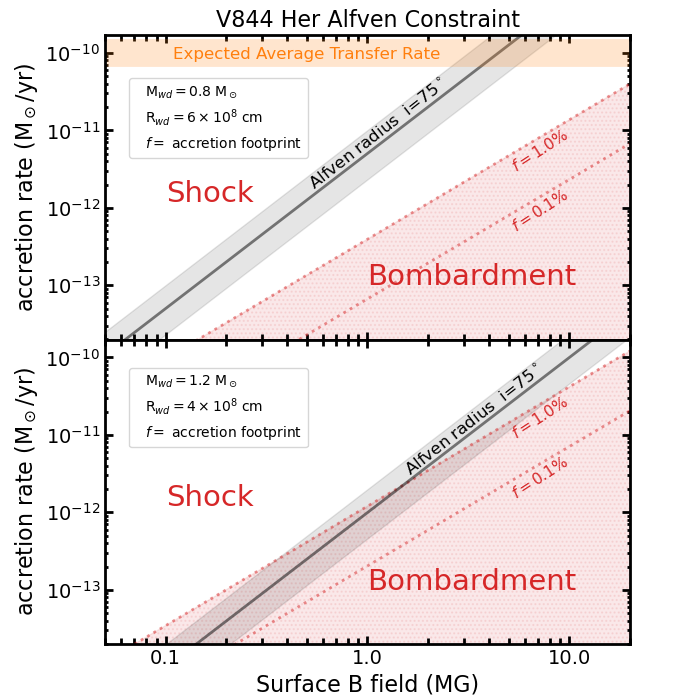}
    \caption{{\bf Top:} The constraints on the surface magnetic field and WD accretion rate (gray band) based on the estimated Alf\'ven radius and assuming a WD mass of 0.8~M$_\odot$. The width of gray band corresponds to the observed uncertainty in the truncation velocity. Assuming a lower orbital inclination moves the constraint to higher accretion rates. The bombardment regime is delineated by the red shaded region. The accretion footprint, $f$, is the fraction of the WD surface receiving the accreting gas. {\bf Bottom:} Same as above, but for a WD mass of 1.2~M$_\odot$. For a high mass WD, the system may reach the bombardment regime during quiescence. 
    \label{bombardment}}
\end{figure}

We use V844~Her to illustrate this by extending the \citet{busschaert15} calculations to IP-like magnetic field strengths, and we plot the result in Figure~\ref{bombardment}.

The orbital period of V844~Her is near the period minimum implying an average mass transfer rate of $< 10^{-10}$~M$_\odot$yr$^{-1}$ \citep{knigge11}. While low, this rate is still sufficient to form an accretion shock at the magnetic field strengths expected in IPs. Between superoutbursts the actual accretion rate onto the WD must be significantly lower than this average \citep{godon06}. However, the quiescent accretion rate onto WDs in superoutbursting systems near the period minimum is not well constrained. 

To place constraints the combination of accretion rate and magnetic field for V844~Her, we estimated the disk truncation radius from the highest disk velocities seen in the $H\alpha$ emission. From that, we derived the Alf\'ven radius using the prescription in \citet{campbell11}. We equated that to the Alf\'ven surface calculated by \citet{lamb73} for radial accretion onto magnetized neutron stars and modified by \citet{ghosh79} to account for accretion from a Keplerian disk. We then solved for the accretion rate as a function of WD surface magnetic field strength, leaving the free parameters of WD mass, $M_{wd}$; WD radius, $R_{wd}$; orbital inclination, $i$; and the accretion footprint as a fraction of the WD surface, $f$.

The resulting constraints on the WD accretion rate as a function of the surface magnetic field for V844~Her are shown by the diagonal gray bands in Figure~\ref{bombardment}.  For a WD with a mass of $0.8$~M$_\odot$ and $B=1$~MG, the quiescent accretion rate is expected to be about $5\times10^{-12}$~M$_\odot$~yr$^{-1}$. The WD is clearly in the shock regime for IP-like surface field strengths. Only for a fairly massive WD does the Alf\'ven radius constraint place the quiescent accretion rate near the border of bombardment regime. 

We conclude that in quiescence, V844~Her is not in the bombardment regime unless it possesses a very massive WD, or has a magnetic field larger than expected for an IP.

\end{appendix}

\end{document}